# Analysis of Diffusion of Ras2 in *Saccharomyces cerevisiae* Using Fluorescence Recovery after Photobleaching

Running title: Diffusion of Ras2 in yeast

Kalyan C. Vinnakota[1, 3], David A. Mitchell[2, 4], Robert J. Deschenes[2, 4], Tetsuro Wakatsuki[1, 3] and Daniel A. Beard[1, 3]

Biotechnology and Bioengineering Center[1], Biochemistry[2], Physiology[3], Medical College of Wisconsin, Milwaukee, Wisconsin 53226, USA
Molecular Medicine[4], College of Medicine, University of South Florida, Tampa, Florida 33612, USA

Corresponding author:
Daniel A. Beard
Department of Physiology,
Medical College of Wisconsin,
8701 Watertown Plank Road,
Milwaukee, WI 53226

E-mail: dbeard@mcw.edu

# Abstract

Binding, lateral diffusion and exchange are fundamental dynamic processes involved in protein association with cellular membranes. In this study, we developed numerical simulations of lateral diffusion and exchange of fluorophores in membranes with arbitrary bleach geometry and exchange of the membrane localized fluorophore with the cytosol during Fluorescence Recovery after Photobleaching (FRAP) experiments. The model simulations were used to design FRAP experiments with varying bleach region sizes on plasma-membrane localized wild type GFP-Ras2 with a dual lipid anchor and mutant GFP-Ras2C318S with a single lipid anchor in live yeast cells to investigate diffusional mobility and the presence of any exchange processes operating in the time scale of our experiments. Model parameters estimated using data from FRAP experiments with a 1 µm × 1 µm bleach region-of-interest (ROI) and a 0.5 µm × 0.5 µm bleach ROI showed that GFP-Ras2, single or dual lipid modified, diffuses as single species with no evidence of exchange with a cytoplasmic pool. This is the first report of Ras2 mobility in yeast plasma membrane. The methods developed in this study are generally applicable for studying diffusion and exchange of membrane associated fluorophores using FRAP on commercial confocal laser scanning microscopes.

# Introduction

Fluorescence recovery after photobleaching (FRAP) has been widely used as a method to measure mobility of fluorophores in solutions and within live cells [1]. Understanding processes such as diffusion, exchange and binding, which determine the recovery of fluorescence during a FRAP experiment, is critical for elucidating the complex nature of the interactions between membrane associated proteins in a lipid environment. Commercially available confocal laser scanning microscopes (CLSM) and software packages have made the capability to perform FRAP widely accessible. However, confocal FRAP data are more difficult to interpret in comparison to Gaussian spot photo-bleaching due to the complexity in taking into account the laser scanning process [2]. In addition to lateral diffusion, chemical exchange processes can also contribute to the recovery of fluorescence in FRAP experiments when extrinsic membrane proteins are being considered [2, 3].

Prior work on interpretation of confocal FRAP data for estimating mobilities of fluorophores in solutions and in cells was focused on numerical simulations for line FRAP or derivation of analytical solutions for FRAP on 2-D ROI's [4-9]. FRAP in systems with diffusion and binding to immobile binding sites has been analyzed by Sprague et al. [9], and later generalized by Dushek et al. [10] for binding to mobile sites localized to a membrane. To our knowledge, there is no known simulation in the literature for diffusion and exchange of membrane associated fluorophores with cytosol, which accounts for confocal point spread function and recovery during the bleach frame. The effects of bleaching and scanning beam point spread functions are significant when the beam waist is comparable to the size of the bleach region of interest, especially in a small region of plasma membrane analyzed in this study. In the present study we developed computational models to study the mobility and exchange of membrane associated proteins in the yeast, *Saccharomyces cerevisiae,* using FRAP. Specifically, we developed computational models describing lateral diffusion of fluorophores within the plasma membrane and exchange of fluorophores between plasma membrane and the cytosol to design our FRAP experiments and interpret our data on the fluorescence recovery of GFP-tagged yeast Ras2 GTPase (GFP-Ras2). Ras proteins are evolutionarily conserved, membrane-associated, GTP binding proteins involved in several signal transduction pathways. Mutations in *ras* genes have been implicated in over 30% of all human cancers. Like its mammalian counterparts, yeast Ras2 protein is initially synthesized as a soluble precursor that associates with the cytosolic face of the plasma membrane only after posttranslational lipidation of the C-terminus of the protein with farnesyl and palmitate [11-13]. Although the use of FRAP to study the movement of membrane associated proteins in mammalian cells is not novel [14], the use of confocal FRAP to measure diffusion and membrane exchange phenomena on the spatial scale and spherical surfaces of yeast membranes presents unique challenges that we have addressed here.

The computational models for diffusion and exchange processes were tested against data from FRAP experiments with different bleach ROI sizes. The ROI size analysis was designed to detect any exchange of GFP-Ras2 in addition to diffusion. Henis et al. [3] developed a similar technique using Gaussian spot photobleaching by changing beam size to determine the presence of exchange processes in addition to diffusion. The development of ROI size analysis using confocal FRAP and numerical simulations of lateral diffusion and membrane-to-cytosol exchange for data analysis opens up new opportunities for membrane studies in the genetically tractable yeast system.

# Methods

## Microbiological techniques

Yeast and bacterial strains were propagated by standard methods [15]. YEPD (1% Yeast Extract, 2% Bactopeptone and 2% glucose) and Synthetic Dextrose (SD) media supplemented with amino acids were prepared as described previously [16]. Bacterial transformations, DNA preparations, and plasmid constructions were performed by standard methods [15]. Yeast transformations were performed by the $Li^+$ ion method [17].

### *Yeast Strains and Plasmids*

The yeast strains used in this study were isogenic to the parental strain BY4742 (Open Labs, Inc.) and have the genotype *MATa his3Δ1 leu2Δ0 lys2Δ0 ura3Δ0*. Construction of B1510 (*GFP-RAS2*), B861 (GFP-Ras2C318S) and B1456 (GFP-PMA1) have been described previously [18, 19]. All GFP-Ras2 allele plasmids contain the Enhanced Green Fluorescent Protein (GFP S65G S72A) fused to the *RAS2* open reading frame under the control of the MET25 promoter. GFP-PMA1 is driven by the GAL1,10 promoter. In experiments utilizing FRAP, cells were grown in synthetic dextrose medium lacking leucine (for selection) and methionine (for maximum fusion protein expression). GFP-PMA1 expression was induced with galactose (2%) in synthetic medium lacking leucine.

### *Yeast cell and glass slide preparation*

Glass slides were prepared by coating wells with 5µg of Concanavalin A (Canavalia ensiformis, Jackbean; US Biologicals, Inc.) in water. The treatment was allowed to dry for 15 minutes at room temperature. Yeast strains harboring GFP-RAS2, GFP-Ras2C318S or GFP-PMA1 were grown in SD medium with the appropriate supplements to an optical density (OD600) of between 0.8 and 1.2. Approximately $4x10^5$ cells (20µl) were spotted onto the concanavalin A treated wells and allowed to adhere for 10 minutes in the dark. Non-adhering cells were removed with repeated washes of water (4x 20µl). Mounting medium (YEPD+1% low-melting agarose (Top Vision LM GQ, Fermentas)) was melted and cooled to room temperature. Approximately 200µl of mounting medium was placed onto the slide. The cover slip was placed on top of the medium and the excess medium and air bubbles removed by applying pressure to the cover slip. The mounting medium was allowed to solidify at room temperature for 15 minutes.

## Confocal microscopy

### *Acquisition of FRAP data*

FRAP experiments were performed with a LEICA TCS SP5 confocal laser scanning microscope using a 63x 1.2 numerical aperture (NA) apochromat water objective. The GFP tagged proteins were excited with a 488 nm laser scanning at 1400 lines per second in bidirectional mode and the fluorescence emission between 500 nm and 550 nm was recorded. Images were acquired at approximately 133 ms per frame. FRAP was performed on the plasma membrane associated pool of GFP-Ras2 proteins by imaging the cell surface with the laser beam incident perpendicular to the cell surface. The curvature of yeast cell determines the upper limit of the ROI to not exceed 2 µm × 2 µm, which will be referred to as the Imaged Area (IA) in this manuscript. Square ROI's of 1 µm × 1 µm and 0.5 µm × 0.5 µm were bleached during FRAP experiments. The confocal microscope was operated in fly-mode, where the scanning beam bleaches during the forward sweep with intensity modulation for bleaching in the ROI and records the signal with scanning intensity during the fly-back. This process ensures a near-instantaneous recording of the post-bleach intensity for every bleached line. We also performed FRAP on GFP-PMA1, an integral membrane protein with at least ten membrane spanning domains, as an internal control based on a significantly lower mobility of integral membrane proteins when compared to lipid anchored proteins such as Ras2 [20]. The FRAP data were normalized to account for fluorophore loss during scanning, which were then analyzed by numerically solving reaction-diffusion equations in time and two spatial dimensions.

## Data Analysis and Computational modeling of FRAP

*FRAP data normalization and model assumptions*

The mean fluorescence intensity in the bleach region of interest (ROI) was normalized with respect to the initial fluorescence and to the imaged area fluorescence (IA) with background subtraction (BG) [14, 21]:

$$F(t) = 100 \frac{[F_{ROI}(t) - F_{BG}(t)]}{[F_{ROI}(0) - F_{BG}(0)]} \frac{[F_{IA}(0) - F_{BG}(0)]}{[F_{IA}(t) - F_{BG}(t)]} \tag{1}$$

The normalization in Eq. 1 accounts for the fluorescence loss due to bleaching during imaging. Since the normalized data account for fluorophore loss due to scanning, the computational models for bleaching and recovery do not include fluorophore loss due to scanning.

We tested two hypotheses expressed in two models against the data to determine the processes contributing to fluorescence recovery during the FRAP experiments. *Model 1* treats the GFP-Ras2 as a single species diffusing laterally in the membrane and *Model 2* tests whether the FRAP data can be explained by a single species that diffuses in the membrane and exchanges with the cytosol.

*Model 1: Diffusion in 2 dimensions*

Concentration changes in fluorescent GFP-Ras2 due to diffusion and bleaching by laser scanning are simulated by the following equation taking into account the bleach ROI geometry, and the beam point spread function,

$$\frac{\partial}{\partial t} C(x,y,t) = D\nabla^2 C(x,y,t) - \alpha K_n(x,y) C(x,y,t) \;,\; 0 \le t \le T_b, \tag{2}$$

where $C$ is the concentration of the fluorophore, $D$ is the diffusivity, $\alpha$ is the laser illumination dependent bleach rate constant and $K_n(x, y)$ is the normalized bleaching light distribution and $T_b$ is the duration of the bleaching frame. Eq. 2 assumes that the fluorescence signal is directly proportional to the concentration of the fluorophore. This equation was solved numerically using a Crank-Nicholson scheme with periodic boundary conditions [22]. Details of the implicit Crank-Nicholson scheme are given in the Appendix along with computer code written in Matlab® (The MathWorks Inc., Natick MA).

To simulate bleaching, the bleaching light distribution $K(x, y)$ is given by the convolution of the ROI shape function $B(x, y)$ and the bleaching beam intensity $I_b(x, y)$.

$$K(x, y) = B(x, y) \otimes I_b(x, y) \tag{3}$$

The normalized bleaching light distribution is defined as:

$$K_n(x, y) = \frac{K(x, y)}{K(0,0)} \tag{4}$$

Following Braeckmans et al.[6], we define the ROI shape function as:

$$B(x, y) = 1, -l/2 \leq x, y \leq l/2$$
$$B(x, y) = 0, l/2 < |x|, |y| \tag{5}$$

The interline distance in our experiments (0.016 μm) is much smaller than the $1/e^2$ radius $w_0$ (0.248 μm) of the bleaching beam, which allows for treating the bleaching light distribution as the convolution of the square bleach region $B(x, y)$ with the bleaching beam intensity $I_b(x, y)$. If the interline distance is much larger than $w_0$, bleaching process should be simulated line-by-line [6].

$I_b$ is the intensity of the laser beam during bleaching.

$$I_b(x, y) = I_{0,b} \exp\left[-2\left(\frac{x^2 + y^2}{w_0^2}\right)\right], \tag{6}$$

where $I_{0,b}$ is the intensity at (0, 0) and $w_0$ is the $1/e^2$ radius of the bleaching beam. $w_0$ is approximated as $0.61\lambda_{ex}/NA$, where excitation wavelength $\lambda_{ex}$ = 488 nm and the numerical aperture NA = 1.2 [23]. During numerical computations, $I_b(x,y)$, was generated for a square grid that was $3w_0 \times 3w_0$ in size and centered at (0, 0). The $1/e^2$ radius of the confocal imaging point spread function $w_{s,0}$ is related to $w_0$ and the excitation and emission wavelengths for an infinitesimally small pinhole by the following equation [23]:

$$w_{s,0} = \frac{w_0}{\sqrt{1 + (\lambda_{ex}/\lambda_{em})^2}}, \tag{7}$$

where $\lambda_{em}$ is the emission wavelength of the fluorophore.

The data were normalized to account for fluorophore loss due to scanning during the post-bleach recovery phase. Therefore, the post-bleach recovery curves were analyzed by the following 2-D diffusion equation without any bleaching term.

$$\frac{\partial}{\partial t} C(x, y, t) = D \nabla^2 C(x, y, t), \quad t > T_b. \tag{8}$$

$C(x, y, t)$ obtained by solving Eqs. (2) and (8), was translated into the simulated fluorescence signal by performing the same normalization used for the fluorescence data in Eq. 1:

$$F_{simulated}(t) = \frac{\langle C_{ROI}(x, y, t) \rangle \langle C_{IA}(x, y, 0) \rangle}{\langle C_{ROI}(x, y, 0) \rangle \langle C_{IA}(x, y, t) \rangle}, \tag{9}$$

where $C_{ROI}(x, y, t)$ is computed as the product of $C(x, y, t)$ and the normalized convolution of the bleach shape function $B(x,y)$ and a 2-D Gaussian imaging point spread function whose $1/e^2$ radius was defined in Eq. 7. Brackets $\langle \cdot \rangle$ indicate averaging over the membrane area.

Diffusivity $D$ and the bleaching rate constant $\alpha$ were estimated for GFP-Ras2 by fitting the $F_{simulated}(t)$ to averaged data from 1 μm × 1 μm bleach ROI and 0.5 μm × 0.5 μm bleach ROI experimental data. The parameter estimation was performed using the fminsearch function in Matlab R2009a using a weighted residual sum of squares (WRSS) as the cost function. Parameter precisions expressed as percent coefficients of variation (CV) were estimated by computing an approximate variance-covariance matrix as detailed in Appendix C. Akaike Information Criterion (AIC) and Schwarz Criterion (SC) were calculated for determining model parsimony. The estimated parameters and their percent CV are reported in Table 1 and the parsimony criteria in Table 2.

*Model 2: Diffusion in 2 dimensions with exchange between the 2-dimensional membrane and a well mixed volume representing the cytosol*

To test the hypothesis that GFP-Ras2 diffuses in the membrane and exchanges with the cytosol where it diffuses faster, we solved a diffusion-exchange model described here. Our model is based on mass balance of total fluorophore given by the following equation:

$$C_c V_c + \langle C_m \rangle A_m = M, \qquad (10)$$

where $C_c$ is the concentration of the fluorophore in the cytosol, which is assumed spatially homogenous; $C_m$ is the concentration of the fluorophore in the membrane; $V_c$ is the volume of cytosol; $A_m$ is the area of the plasma membrane; and $M$ is the number of moles of the fluorophore.

The insertion and removal of GFP-Ras2 are modeled as first order reaction:

$$\frac{\partial C_m}{\partial t} = D_m \nabla^2 C_m - k_{off} C_m + k_{on} C_c. \qquad (11)$$

Substituting Eq. 10 into Eq. 11:

$$\frac{\partial C_m}{\partial t} = D_m \nabla^2 C_m - k_{off} C_m + k_{on} \left( \frac{M}{V_c} - \frac{A_m}{V_c} \langle C_m \rangle \right), \; t > T_b. \qquad (12)$$

Diffusion, bleaching and exchange are accounted for during the bleach phase:

$$\frac{\partial}{\partial t} C(x,y,t) = D \nabla^2 C(x,y,t) - k_{off} C_m + k_{on} \left( \frac{M}{V_c} - \frac{A_m}{V_c} \langle C_m \rangle \right) - \alpha K_n(x,y) C(x,y,t), \; 0 \leq t \leq T_b, \qquad (13)$$

Eqs. 12 and 13 are solved numerically by solving the diffusion process using the Crank-Nicholson scheme developed for *Model 1* and the exchange process sequentially during each time step by means of operator splitting. These methods are detailed in Appendices A and B. For the purpose of simulating diffusion and exchange processes, the volume of an average yeast cell is approximated by that of a sphere of 5 micron diameter and the area of the square membrane domain was set equal to the surface area of this sphere. From preliminary Fluorescence Correlation Spectroscopy (data not shown) measurements of particle density per micron squared, we calculated the concentration of the fluorophore on the membrane to be 0.01 nmol/dm$^2$ or 10 pmol/dm$^2$. In addition, we assumed that the ratio of number of fluorophores in the membrane to that in the cytosol was 1000:1, resulting in a nanomolar cytosolic concentration of the fluorophore. The relationship between number of fluorophores in the cytosol and the membrane leads to an equilibrium relationship between $k_{on}$ and $k_{off}$:

$$\frac{k_{on}}{k_{off}} = \frac{C_m}{C_c} = 1000 \frac{V_c}{A_m} = 8.33 \times 10^{-3} \text{ dm}. \qquad (14)$$

Eq. 14 provides a constraint so only one rate constant may be treated as an adjustable parameter. We treat $k_{off}$ as an adjustable parameter during parameter estimation and model simulations. The volume of distribution of cytosolic proteins will be the water space of cytosol excluding the spaces occupied by other organelles. Treating the entire intracellular space as the volume of distribution for the cytosolic Ras2 may result in an underestimation of cytosolic concentration of the fluorophore. Only a detailed study using data from morphometry and cell composition can address this question [24], which is beyond the scope of the present study. Changes in volume estimates will be reflected in the rate constants and concentration estimates, which will be lumped in the parameter estimates for the rate constants in the present study.

FRAP simulations for both diffusion and exchange are performed by computing diffusion and exchange solutions sequentially for each time step. Statistical analyses were performed on parameter estimates were performed using methods detailed in Appendix C.

## Results and Discussion

*Subcellular localization and photobleaching of GFP-RAS proteins in yeast*
The sub-cellular localization of GFP-Ras2 (double lipidated) and GFP-Ras2C318S (single lipidated) was examined in wild type yeast cells. As previously observed, GFP-Ras2 is primarily localized on the inner leaflet of the plasma membrane (Fig. 1A). In contrast, the sub-cellular localization of singly lipidated Ras2 is distributed between endomembranes and the plasma membrane, with the majority of the protein on endomembranes (70%) (Fig. 1B). However, the pool of GFP-Ras2C318S on the plasma membrane is sufficient for the FRAP analysis. Fig. 2 shows an example of a cell in which a 1 μm × 1 μm ROI has been photobleached and the recovery of fluorescence measured. Images are collected at ~133 ms/frame and continuing for 20 seconds until complete recovery was observed. The normalized FRAP data showed complete recovery for all strains imaged. Therefore we did not include an immobile fraction in modeling the data, which would have shown up as a non recoverable fraction in the data. While collecting FRAP data from cells expressing GFP-Ras2-C318S, we note that any protein present in membranous structures within the ROI under the plasma membrane could be bleached. However, the experiment design is still capable of answering the question whether the recovery of fluorescence is aided by plasma membrane localized fluorophores in the region surrounding the bleached ROI that exchange with cytosol where they may diffuse faster and exchange with the plasma membrane in the bleached ROI.

*Analysis of diffusion of wild-type and non-palmitoylated GFP-Ras2 by FRAP*
Fluorescence recovery data were collected from cells expressing GFP-Ras2 and GFP-Ras2C318S (non-palmitoylated) at 1 μm × 1 μm and 0.5 μm × 0.5 μm bleach ROI sizes. The normalized fluorescence recovery data for GFP-Ras2 were analyzed by the two models described in the Methods section. *Model 1* accounts for single-species diffusion in the cell membrane, while model 2 accounts for both diffusion and exchange of protein between the membrane and the cytosol.

Fig. 3A shows FRAP recovery data and model fits for both ROI sizes for GFP-Ras2. The FRAP data for the smaller bleach ROI size shows a faster recovery when compared to the larger ROI data, which would be expected for a diffusion coupled phenomenon i.e., the recovery is not entirely governed by exchange with cytosol. Note that the depth of photobleaching for the smaller ROI turned out to be greater than that of the larger ROI. Model predictions of FRAP recovery for a given diffusivity and bleach rate constant shows that the depth of photobleaching of the smaller ROI cannot be greater than that of the larger ROI (Fig 4A). Parameters of *Model 1* and *Model 2* for GFP-Ras2 were estimated both with and without the data for the bleach frame and the immediate postbleach frame. Estimated parameters listed in Tables 1 and 2 show that the parameter estimates ($D$ and $\alpha$) for *Model 1* for the larger ROI vary very little with inclusion or exclusion of bleach frame during fitting. Parameter estimates for the smaller ROI show that the inclusion of the bleach frame during fitting results in higher estimates of $D$ and $\alpha$, but exclusion of the bleach frame during fitting brings these estimates very close to and statistically not different from the larger ROI at 95% significance level. In addition, the residuals from the fits that excluded the bleach frame and the immediate postbleach frame showed an autocorrelation that could be expected from a random signal at 95% significance level whereas the residuals from the fit that included all frames showed a more structured correlation indicating a slightly poorer fit. (Differences between these fits re not apparent to the naked eye when plotted on the axes scale of Figure 3.) Mueller et al.[25] found that detector blinding or possible transient loss in postbleach detector sensitivity following saturating laser intensity detection in the bleach frame could result in erroneous detection of fluorescence in the immediate post bleach frames and consequently very high estimates of diffusivity for large bleach ROI's. Since the depth of photobleaching discrepancy was found in the smaller ROI data for GFP-Ras2, and not found in all datasets as demonstrated by parameter estimation results with and without bleach frame data, we may not attribute it to detector blinding.

Parameter estimates for *Model 2* show that the additional parameter $k_{off}$ is essentially indistinguishable from zero for all cases considered and the Akaike and Schwarz criteria, listed in Table 2, show that *Model 1* is preferable to *Model 2*. The weighted residual sum of squares for *Model 2* fit of 1 μm × 1 μm ROI data show marginal improvement. However, AIC, SC and the P-Value of F-test show that *Model 1* is preferable to *Model 2*. In summary, we may conclude that the motion of GFP-Ras2 in the time scale of our experiments is well described by diffusion and that mean diffusivity estimate for GFP-Ras2 is ~0.07 $\mu m^2$/s.

Fig. 3B shows FRAP recovery data and model fits for both ROI sizes for GFP-Ras2C318S, which is not palmitoylated. The FRAP recovery data for the smaller bleach ROI size recovers faster and has a smaller depth of photobleaching when compared to the recovery data from the larger bleach ROI size, which is consistent with a diffusion coupled recovery (Figure 4 B). The parameter estimates pertaining to *Model 1* listed in Table 1 show that the estimates of D and a for the non-palmitoylated GFP-Ras2C318S are very close to each other and statistically indistinguishable at 95% significance level. Model 2 parameter estimates show that the additional parameter $k_{off}$ is very close to and indistinguishable from zero. The WRSS for *Model 2* is higher than that for *Model 1* and Akaike and Schwarz criteria are lower for *Model 1*, which makes it preferable to *Model 2*. Therefore, GFP-Ras2C318S FRAP recovery is governed by diffusion with a mean diffusivity ~0.28 $\mu m^2$/s.

Fig. 5A and 5B show the simulations of FRAP *Model 2* for wild-type GFP-Ras2, where membrane bound fluorophore exchanges with a well mixed cytosolic compartment. The bleach ROI dependence of recovery solutions for 1 μm × 1 μm and 0.5 μm × 0.5 μm ROI sizes varies with the rate of exchange. A fast exchange rate makes the FRAP recovery independent of bleach ROI size as expected in a reaction dominated process (Fig 4B) [3, 9]. Smaller rate constants representing slower exchange also show faster recovery when compared to recovery from pure lateral diffusion (Fig 4A). Figures 5 C and D demonstrate diffusion-exchange model simulations for GFP-Ras2C318S data, confirming the findings in Fig. 5 A and B. Clearly, *Model 2* can match the data as well as Model 1, but does so only when it actually does reduce to *Model 1*. Therefore *Model 2* may indeed effectively describe what is happening. However, it does so in such a way that the exchange processes do not occur with observable significance over the timescales of these experiments.

Several conclusions can be made from these data. First, single and dual lipidated forms of GFP-Ras2 exhibit different rates of diffusion when associated with the yeast plasma membrane (Table 1). The single modified form, which only has a farnesyl moiety, diffuses approximately 4 times faster than the dual lipidated form containing both farnesyl and palmitoyl modifications. Mammalian H-Ras undergoes the same two modifications, but the estimated diffusivity of Ras2 in yeast plasma membrane (0.075 $\mu m^2/s$) is slower than that of H-Ras (1 $\mu m^2/s$ at 22 ºC) and K-Ras (1 $\mu m^2/s$ at 37 ºC or 0.63 $\mu m^2/s$ at 22 ºC) observed in mammalian cell plasma membranes [26, 27]. The slower diffusivity of Ras2 in yeast plasma membrane could be due to different membrane composition of yeast membranes compared to mammalian plasma membranes. Henis et al. [3] have shown that Ras diffusion is sensitive to lipid composition of the plasma membrane. We also performed FRAP on a GFP-fused integral membrane protein, PMA1 (GFP-PMA1). GFP-PMA1, an integral membrane protein with at least ten membrane spanning regions, diffuses much more slowly than a lipid anchored protein such as Ras2. Fig. 6 shows normalized fluorescence recovery of both GFP-Ras2 and GFP-PMA1 for 1 $\mu m \times$ 1 $\mu m$ ROI. GFP-PMA1 normalized fluorescence recovers fully at approximately 250 seconds, whereas its non-normalized fluorescence recovers to only 40 percent of the initial fluorescence (data not shown). This result is in agreement with the study by Valdez-Taubas and Pelham on another trans-membrane integral protein Sso1 [28].

Unexpectedly, the fluorescence recovery of both forms GFP-Ras2 (single and double lipidated) in yeast is dominated by diffusion, with no evidence for exchange. Experiments designed to measure exchange of lipidated peptides between model membranes have shown that dual lipidated (prenyl and palmitate) peptides exchange slowly with intervesicle equilibration half-times in hours [29], whereas single lipidated peptides (farnesyl) exchange very rapidly with intervesicle equilibration half time in seconds [30]. However, this is not observed for the singly farnesylated GFP-Ras2C318S proteins. There are of course many differences between model membranes and cells with respect to sterol-phospholipid stoichiometries and the presence of high protein concentrations in cell membranes [31]. It might also reflect the difference between the composition of yeast and mammalian plasma membranes. The lipid and protein composition of mammalian and yeast plasma membrane differs greatly [31-33] with the role of lipid composition of the membrane exerting a major influence on whether the Ras mobility is diffusion or exchange dominated [3]. Henis et al. [3] have demonstrated that cholesterol depletion in COS-7 cells changes the mobility of palmitoylation mutants of H-Ras from being dominated by lateral diffusion to exchange dominated behavior. The diffusivity of these mutants was measured to be 5 $\mu m^2/s$ at 22 ºC, which is five fold higher than that of H-Ras [34]. Goodwin et al. [34] concluded that a significant fraction of H-Ras palmitoylation mutant existed as a soluble cytoplasmic fraction based on FCS data however their data were not tested with FRAP beam size or ROI size analyses.

Finally, we noted that the normalized fluorescence of GFP-Ras2 recovers as a single pool of GFP-Ras2 implying that the fluorophore diffuses as a single species. This was also somewhat surprising because of the nature of the post translational modifications at the C-terminus of the Ras2 protein. Ras2 was chosen for this study because it is initially synthesized as a soluble precursor and rapidly associates with the cytosolic face of the plasma membrane through dual lipidation (farnesylation and palmitoylation) of the C-terminal CC*aaX* box [11-13]. Farnesylation is irreversible, whereas palmitoylation is potentially reversible and and hence, Ras has the potential to cycle between the acylated and deacylated state. Based on this notion, the acylation/deacylation cycle is proposed to be required for proper subcellular trafficking of Ras2 [27, 35]. The recovery of GFP-Ras2 fluorescence as a single species in the wild type strain suggesting that Ras2 is almost entirely dually lipidated. Any significant presence of the non-palmitoylated form of Ras2 would have to be described by a two-species diffusion model, which would result in a failure of the single species model. The absence of exchange suggests that acylation does not "lock" the molecule into membrane association. Therefore, one possible model is that palmitoylation directs GFP-Ras2 to the plasma membrane. Consistent with this, the distribution of nonpalmitoylated GFP-Ras2C318S in cells at "steady-state" is primarily on endomembranes (Fig. 1B).

*Confocal FRAP*
Commercial confocal microscopes have made FRAP widely accessible. However, quantitative analysis of confocal FRAP requires development of appropriate mathematical models pertinent to the experiment and the system being studied. The presence of chemical reactions complicates the analysis even in the case of Gaussian spot photobleaching where the theory is well developed [2]. Peters and colleagues [4, 5] developed line-scanning microphotolysis and numerical simulations to measure lateral and 3D diffusion in solutions and in cells. Braeckmans et al. [6, 36] described the mathematical basis for 3D FRAP with an arbitrary 2D geometry bleach ROI and for line FRAP, and Braga et al.[8] developed a model that accounts for diffusion of fluorophores during bleaching. These efforts were focused on deriving analytical expressions that can be used to fit data to facilitate ease of use by biologists in general. A particular feature of such approaches is the description of postbleach intensity using an analytical function with estimable parameters that must be fitted to the postbleach intensity profile. This is critical since this fitted analytical solution constitutes the initial condition for the recovery. In contrast, our approach has only one initial condition, which is the prebleach fluophore concentration distribution. The bleach frame and recovery are computed by simulating the bleaching and recovery processes with the same parameters for the entire model, requiring no additional fitting of the bleach profile alone. Fig 7 shows that the *Model 1* predictions of the spatial bleach frame intensity profile are validated by experimental image data. We also account for the change in total fluorophore concentration during bleaching to account for fluorophore mass conservation, which is crucial for simulating the exchange of fluorophore in *Model 2*. Sprague et al. [9] developed models to account for diffusion and binding to immobile binding sites in circular spot photobleaching experiments. Dushek et al. [10] theoretically analyzed a membrane localized ligand-receptor binding and diffusion problem where the binding sites are not immobile to determine parameter regimes for model identifiability.

In our present study we address a different biophysical problem, where the membrane associated fluorophores, which diffuse fast in the cytosol, upon association with the membrane, diffuse at a slower rate and are not immobile. Our simulations account for both translational diffusion and exchange processes, and diffusion during bleaching. FRAP experiments were performed with different bleach ROI sizes to distinguish between possible mechanisms contributing to the recovery process. A similar experimental technique was developed by Henis et al. [3] using Gaussian-spot FRAP by changing the beam size to determine the contribution of chemical exchange processes to FRAP recovery in addition to translational diffusion of Ras isoforms in mammalian cells. Experimental data reported by Henis et al. [3] indicate that the exchange of free cytosolic fluorophore is not diffusion-limited, which is consistent with our simulation approach. Henis et al. [3] point to the difficulties in performing and interpreting confocal FRAP for analyzing diffusion and exchange processes, such as time of photobleaching and image acquisition, which were overcome partly in our study with small bleach ROI's and fast image acquisition combined with data analysis by computational modeling. In summary we were able to quantitatively test whether wild type GFP-Ras2 and palmitoylation mutant GFP-Ras2C318S in yeast diffused laterally in the plasma membrane as a single species on the time scale of our FRAP experiments.

*Conclusions*
ROI size analysis of FRAP experiments shows that the data are consistent with a single-species lateral diffusion model. Simulations of diffusion and exchange of GFP-Ras2 between membrane and cytosol show that the spatial dependence of FRAP curves can be used to account for exchange if it occurs. Larger rates of exchange diminish the spatial dependence of FRAP. The singly lipidated mutant GFP-Ras2C318S also diffuses as a single species in the plasma membrane. In summary the present study developed a confocal microscopy based method to investigate the mobility of membrane associated fluorophores and reported for the first time the mobility and the nature of diffusion of wild type GFP-Ras2 and singly lipidated mutant GFP-Ras2C318S in the yeast, *Saccharomyces cerevisia*e.

# Appendix

## A. Simulation of diffusion equation with bleaching

The 2-D diffusion process governed by Eq. 8 was approximated with a finite difference scheme using the Alternating Direction Implicit (ADI) method in the following manner:

$$\frac{C^*(x,y,t+\delta t) - C(x,y,t)}{\delta t} = \frac{D}{2}\left[\frac{C^*(x+\delta x,y,t) - 2C(x,y,t) + C^*(x-\delta x,y,t)}{\delta x^2} + \frac{C^*(x+\delta x,y,t+\delta t) - 2C(x,y,t+\delta t) + C^*(x-\delta x,y,t+\delta t)}{\delta x^2}\right], \quad (A1)$$

$$\frac{C(x,y,t+\delta t) - C^*(x,y,t+\delta t)}{\delta t} = \frac{D}{2}\left[\frac{C(x,y+\delta y,t) - 2C^*(x,y,t) + C(x,y-\delta y,t)}{\delta y^2} + \frac{C(x,y+\delta y,t+\delta t) - 2C^*(x,y,t+\delta t) + C(x,y-\delta y,t+\delta t)}{\delta y^2}\right], \quad (A2)$$

where $\delta x$, $\delta y$ are the spatial discretization steps and $\delta t$ is the time step.

Eq. A1 was expressed in terms of matrix operators with periodic boundary conditions, where the operator assumes infinite periodic tiling of the simulated region in $x$ and $y$ directions and conserves mass as a consequence:

$$\left[\mathbf{I} - \frac{D\delta t}{2}\mathbf{K}_{diff,x}\right]\mathbf{C}^* = \left[\mathbf{I} + \frac{D\delta t}{2}\mathbf{K}_{diff,x}\right]\mathbf{C}(n), \quad (A3)$$

$$\left[\mathbf{I} - \frac{D\delta t}{2}\mathbf{K}_{diff,y}\right]\mathbf{C}(n+1) = \left[\mathbf{I} + \frac{D\delta t}{2}\mathbf{K}_{diff,y}\right]\mathbf{C}^*, \quad (A4)$$

where, $\mathbf{C}$, $\mathbf{C}^* \in \mathbb{R}^N$, $\mathbf{K}_{diff,x}$, $\mathbf{K}_{diff,y} \in \mathbb{R}^{N \times N}$ are the diffusion operator in $x$ and $y$ directions respectively, $\mathbf{I} \in \mathbb{R}^{N \times N}$ is the identity matrix, $N$ is the number of grid points in the simulation, and $n$ is the time step index.

During bleaching, Eqs. A1 and A2 will have an additional bleaching term on the right hand side:

$$\frac{C^*(x,y,t+\delta t) - C(x,y,t)}{\delta t} = \frac{D}{2}\left[\frac{C^*(x+\delta x,y,t) - 2C(x,y,t) + C^*(x-\delta x,y,t)}{\delta x^2} + \frac{C^*(x+\delta x,y,t+\delta t) - 2C(x,y,t+\delta t) + C^*(x-\delta x,y,t+\delta t)}{\delta x^2}\right] - \quad (A5)$$

$$\alpha K_b(x,y,z)C^*(x,y,t+\delta t/2)$$

$$\frac{C(x,y,t+\delta t)-C^*(x,y,t+\delta t)}{\delta t}=$$

$$\frac{D}{2}\left[\begin{array}{c}\dfrac{C(x,y+\delta y,t)-2C^*(x,y,t)+C(x,y-\delta y,t)}{\delta y^2}+\\ \dfrac{C(x,y+\delta y,t+\delta t)-2C^*(x,y,t+\delta t)+C(x,y-\delta y,t+\delta t)}{\delta y^2}\end{array}\right]- \tag{A6}$$

$$\alpha K_b(x,y,z)C(x,y,t+\delta t)$$

Eq. A2 was expressed in the following manner to include a bleaching operator $\mathbf{K}_r \in \mathbb{R}^{N \times N}$:

$$\left[\mathbf{I}-\frac{D\delta t}{2}\mathbf{K}_{diff,x}+\mathbf{K}_r\right]\mathbf{C}^*=\left[\mathbf{I}+\frac{D\delta t}{2}\mathbf{K}_{diff,x}\right]\mathbf{C}(n), \tag{A7}$$

$$\left[\mathbf{I}-\frac{D\delta t}{2}\mathbf{K}_{diff,y}+\mathbf{K}_r\right]\mathbf{C}(n+1)=\left[\mathbf{I}+\frac{D\delta t}{2}\mathbf{K}_{diff,y}\right]\mathbf{C}^*, \tag{A8}$$

Eqs. A3, A4, A7 and A8 were solved in an iterative scheme simulate bleaching and diffusion processes during FRAP experiments.

## B. Simulation of diffusion and exchange using operator splitting

The fluorophore insertion and removal from the plasma membrane is governed by the following equation:

$$\frac{\partial C_m}{\partial t}=-k_{off}C_m+k_{on}\left(\frac{M}{V_c}-\frac{A_m}{V_c}\langle C_m \rangle\right) \tag{B1}$$

Eq. B1 is expressed in a matrix form for the membrane discretized by a two dimensional grid:

$$\frac{d\mathbf{C}_m}{dt}=\left(-k_{on}\frac{\mathbf{A}_m}{NV_c}-k_{off}\mathbf{I}\right)\mathbf{C}_m+k_{on}\frac{M}{V_c}, \tag{B2}$$

where, $\mathbf{C}_m \in \mathbb{R}^N$, $\mathbf{I} \in \mathbb{R}^{N \times N}$ is the identity matrix, and $N$ is the number of grid points in the simulation.

The eigen values of the matrix multiplying the concentration vector $\mathbf{C}_m$ are:

$$\lambda_1=-k_{on}\frac{A_m}{V_c}-k_{off}, \quad \lambda_2=-k_{off} \tag{B3}$$

The analytical solution for Eq. B2 can be expressed in terms of the eigenvalues in Eq. B3 and the projectors onto the eigenvectors

$$\mathbf{C}_m(t)=\mathbf{P}_1\left(\mathbf{C}_m(0)+k_{on}\frac{M}{V_c\lambda_1}\right)e^{\lambda_1 t}-\mathbf{P}_1 k_{on}\frac{M}{V_c\lambda_1}+\mathbf{P}_2\left(\mathbf{C}_m(0)+k_{on}\frac{M}{V_c\lambda_2}\right)e^{\lambda_2 t}-\mathbf{P}_2 k_{on}\frac{M}{V_c\lambda_2} \tag{B4}$$

The projectors $\mathbf{P}_1$ and $\mathbf{P}_2$ are given by:

$$\mathbf{P}_1 = \frac{1}{N}, \quad \mathbf{P}_2 = \mathbf{I} - \mathbf{P}_1 \tag{B5}$$

Operator splitting is implemented for solving diffusion and exchange during bleaching by first solving Eqs. A7, A8 for one time step followed by Eq. B4 for the same time step. The same scheme is implemented during recovery by first solving Eqs. A3, A4 for one time step followed by Eq. B4 for the same time step. We also update the total mass of the system during the bleach frame simulation to account for loss of fluorophore during photobleaching.

The spatial discretization step size and time step size were chosen by comparing the relative error in the solution with change in spatial and time step sizes, and optimizing for computing time. We find that a spatial step size of 0.05 µm and a time step size (11 ms) gives a solution with less than 5% relative error (relative to sum of squares of residuals at 0.025 µm spatial step size) and the least computing time for that error.

## C. Estimation of parameter precisions and model parsimony criteria

The methods summarized in this section are described in detail by Landaw and Distefano [37]. We follow their terminology and notation in this section of the Appendix. To characterize the precision of our parameter estimates, we calculated an approximate variance-covariance matrix and the percent coefficient of variation of the parameters as per Eqs. 10-12 in Landaw and Distefano [37]. We computed the sensitivity of time courses of fluorescence recovery for a 1% change in the values of the $P$ adjustable parameters $\mathbf{s}(t, \mathbf{p})$, where $\mathbf{p}$ is the vector of P adjustable parameters, $t$ denotes time and $\mathbf{s}$ denotes the vector of sensitivities. For every time point we then computed a $P$-by-$P$ information matrix $\mathbf{M}(t_i)$ using Eq. 10 from Landaw and Distefano (1984) shown below:

$$\mathbf{M}(t_i) = \mathbf{s}(t_i, \mathbf{p})\mathbf{s}^\mathbf{T}(t_i, \mathbf{p}) / \sigma^2(t_i), \tag{C1}$$

where $\sigma^2(t_i)$ is the standard deviation of the experimental measurement at each time point. The total information matrix $\mathbf{M}$ is computed by adding all $\mathbf{M}(t_i)$:

$$\mathbf{M} = \sum_{i=1}^{N} \mathbf{M}(t_i), \tag{C2}$$

where $N$ is the number of time points in the fitted time course.

An approximation to the variance-covariance matrix $\mathbf{COV}(\hat{\mathbf{p}})$ of the optimized values of the adjustable parameters $\hat{\mathbf{p}}$ may be computed by inverting $\sigma^2 \mathbf{M}$, where $\sigma^2$ is treated as 1 since we assume that the error variances are equal to the standard deviations of measured data at each time point. Table 1 reports the computed coefficients of variation ($100 \times \hat{p}_j / \sqrt{COV(\hat{p})_{jj}}$) for the adjustable parameters.

Akaike Information Criterion (*AIC*) and Schwartz Criterion (*SC*) were calculated using the following equations:

$$AIC = WRSS + 2P \tag{C3}$$
$$SC = WRSS + P\ln(N) \tag{C4}$$

To test that any decrease in WRSS with *Model 2*, which has one additional adjustable parameter in comparison to *Model 1*, is statistically significant, we computed the F statistic using the following equation:

$$F = \frac{(WRSS_{Model1} - WRSS_{Model2})}{WRSS_{Model1}/(N - P_2)}, \tag{C5}$$

where $N$ is the total number of points in the fitted time course, $P_2$ is the number of adjustable parameters in *Model 2*. The test statistic has an asymptotic F distribution with (1, $N$-$P_2$) degrees of freedom under the null hypothesis that *Model 1* is correct. A P-value of less than 0.05 will lead to the rejection of the null hypothesis at 95% significance level.

Parameter confidence intervals based on a *t*-distribution were derived to test whether a hypothesized value of a parameter was different in a statistically significant manner from the estimated value at 1-$\alpha$ probability. The number of degrees of freedom of the *t*-distribution for *Model j* is $N$-$P_{Modelj}$.

$$CI = \hat{p}_i \pm t_{1-\alpha/2} SD(\hat{p}_i) \tag{C6}$$

## Acknowledgements

The authors are thankful to Dr. Ranjan Dash for helpful discussions on numerical methods. This work was supported by National Institutes of Health grants HL072011 (DAB) and CA050211 and GM03977 (RJD).

# Tables

Table 1. Estimated Parameters for GFP-Ras2, GFP-Ras2C318S and GFP-PMA1 expressed as mean (% CV)

| Strain | ROI size | Model1 | | Model2 | | |
|---|---|---|---|---|---|---|
| | | $D$ ($\mu m^2/s$) | $\alpha$ ($s^{-1}$) | $D$ ($\mu m^2/s$) | $\alpha$ ($s^{-1}$) | $k_{off}$ ($s^{-1}$) |
| GFP-Ras2 | 1 μm × 1 μm | 0.0691 (3.89) | 27.7 (7.88) | 0.064 (12.98) | 27.28 (9.16) | 0.0214 (136.67) |
| | 0.5 μm × 0.5 μm | 0.0959 (5.75) | 48.2 (11.59) | 0.0957 (16.3) | 47.56 (21.1) | 2.13e-3 (2665.2) |
| GFP-Ras2C318S | 1 μm × 1 μm | 0.2743* (8.95) | 43.09* (24.6) | 0.2863 (9.5) | 49.42 (191) | 3.76e-6 (1.29e7) |
| | 0.5 μm × 0.5 μm | 0.2768* (21.99) | 43.5* (49.41) | 0.2764 (19.8) | 42.85 (621.7) | 3.19e-7 (1.8e9) |
| GFP-PMA1 | 1 μm × 1 μm | 4.34e-4 | 17.55 | - | - | - |

* Cannot reject the null hypothesis that these parameters are not distinguishable from each other between two different ROI sizes for the same strain at 95% significance level

Table 2. Estimated Parameters for GFP-Ras2, GFP-Ras2C318S and GFP-PMA1 expressed as mean (% CV), without including the squares of residuals bleach frame and the immediate postbleach frame intensities in the cost function.

| Strain | ROI size | Model1 | | Model2 | | |
|---|---|---|---|---|---|---|
| | | $D$ ($\mu m^2/s$) | $\alpha$ ($s^{-1}$) | $D$ ($\mu m^2/s$) | $\alpha$ ($s^{-1}$) | $k_{off}$ ($s^{-1}$) |
| GFP-Ras2 | 1 μm × 1 μm | 0.0705* (4.92) | 29.60* (12.26) | 0.0625 (16.47) | 26.95 (14.65) | 0.0240 (137.95) |
| | 0.5 μm × 0.5 μm | 0.0730* (8.39) | 27.88* (14.52) | 0.0976# | 50.07# | 0# |
| GFP-Ras2C318S | 1 μm × 1 μm | 0.3154* (25.73) | 132.53* (190.08) | 0.3154 (54.96) | 130.83 (262) | 2.5e-4 (6.55e4) |
| | 0.5 μm × 0.5 μm | 0.2221* (43.03) | 28.51* (92.12) | 0.2220 (64.35) | 28.17 (870.4) | 2.43e-7 (4.75e9) |
| GFP-PMA1 | 1 μm × 1 μm | 4.34e-4 | 17.55 | - | - | - |

* Cannot reject the null hypothesis that these parameters are not distinguishable from each other between two different ROI sizes for the same strain at 95% significance level

# optimizer hit the lower bound for $k_{off}$

Table 3. WRSS, Akaike Information Criterion (AIC) and Schwarz Criterion (SC) for Models 1 and 2, and P-value for F-test for GFP-Ras2 model fits.

| Strain | ROI size | Model1 | | | Model2 | | | P value of F-test |
|---|---|---|---|---|---|---|---|---|
| | | WRSS | AIC | SC | WRSS | AIC | SC | |
| GFP-Ras2 | 1 μm × 1 μm | 4.33 | 8.3253 | 13.7263 | 4.24 | 10.2375 | 18.3390 | 0.1347 [#] |
| | 1 μm × 1 μm [*] | 4.09 | 8.09 | 13.4543 | 4.01 | 10.0059 | 18.0523 | 0.1507 [#] |
| | 0.5 μm × 0.5 μm | 6.12 | 10.1238 | 15.3341 | 6.16 | 12.1547 | 19.9702 | 1.0000 |
| | 0.5 μm × 0.5 μm [*] | 4.74 | 8.7394 | 13.9093 | 5.8812 | 11.8812 | 19.6361 | 1.0000 |
| GFP-Ras2C318S | 1 μm × 1 μm | 22.29 | 26.2898 | 30.7868 | 24.82 | 30.8182 | 37.5637 | 1.0000 |
| | 1 μm × 1 μm [*] | 20.49 | 24.4900 | 28.9290 | 21.33 | 25.3290 | 33.9875 | 1.0000 |
| | 0.5 μm × 0.5 μm | 5.33 | 9.3321 | 13.5208 | 6.13 | 12.1314 | 18.4145 | 1.0000 |
| | 0.5 μm × 0.5 μm [*] | 4.96 | 8.9600 | 13.0809 | 5.53 | 11.5323 | 17.7137 | 1.0000 |

[*] Model statistics corresponding to fits of the fluorescence recovery without the bleach frame intensity in the cost function

[#] Null hypothesis that *Model 1* (diffusion alone) is correct cannot be rejected at 95% significance level

Figure Legends

1. Localization of GFP-Ras2. Image was acquired using a 63x 1.2 na water objective with high intensity 488 nm excitation and emission collected between 500 nm and 550 nm. A) Wildtype GFP-Ras2 is mainly localized to the plasma membrane B) GFP-Ras2C318S is localized to both plasma and endomembranes.

2. Time-stamped FRAP imaging sequence acquired at 1400 lines/s using bidirectional scanning and fly-mode on a Leica TCS-SP5 CLSM: (A) Pre-bleach image (B) Image acquired during the bleach frame (C) Image acquired 137 ms after the bleach frame (D) Image acquired after 19.8 seconds of recovery. The images are presented in 256 level-inverted grayscale color map, where black represents saturation and white represents zero signal.

3. FRAP data and model fits: (A) GFP-Ras2 FRAP data for 1μm × 1μm (•) and 0.5μm × 0.5μm (•) bleach ROI's, and model fits for 1μm × 1μm (—) bleach ROI and for 0.5μm × 0.5μm (—) bleach ROI yielding the estimated parameters $D = 0.0691$ μm$^2$/s, $\alpha = 27.7$ s$^{-1}$, and $D = 0.0730$ μm$^2$/s, $\alpha = 27.88$ s$^{-1}$ respectively, (B) GFP-Ras2C318S FRAP data for 1μm × 1μm and 0.5μm × 0.5μm bleach ROI's, and model fits for 1μm × 1μm bleach ROI and prediction for 0.5μm × 0.5μm bleach ROI yielding the estimated parameters $D = 0.2743$ μm$^2$/s, $\alpha = 43.09$ s$^{-1}$, and $D = 0.2768$ μm$^2$/s, $\alpha = 43.5$ s$^{-1}$ respectively.

4. FRAP data, model fits and predictions: (A) GFP-Ras2 FRAP data for 1μm × 1μm (•) and 0.5μm × 0.5μm (•) bleach ROI's, and model fit for 1μm × 1μm (—) bleach ROI and model prediction for 0.5μm × 0.5μm (—) bleach ROI using the estimated parameters $D = 0.0691$ μm$^2$/s, $\alpha = 27.7$ s$^{-1}$ for 1μm × 1μm bleach ROI, (B) GFP-Ras2C318S FRAP data for 1μm × 1μm (•) and 0.5μm × 0.5μm (•) bleach ROI's, and model fit for 1μm × 1μm (—) bleach ROI and model prediction for 0.5μm × 0.5μm (—) bleach ROI using the estimated parameters $D = 0.2743$ μm$^2$/s, $\alpha = 43.09$ s$^{-1}$ for 1μm × 1μm bleach ROI

5. FRAP data with model simulations for lateral membrane diffusion and exchange with cytosol.
In each panel, the FRAP data for 1μm × 1μm and 0.5μm × 0.5μm bleach ROI's are plotted using black and red dots, and the corresponding model simulations are plotted using black and red solid lines respectively. (A) GFP-Ras2 data with model predictions for diffusion and slow exchange with parameters $D = 0.0691$ μm$^2$/s; $\alpha = 27.7$ s$^{-1}$; $k_{on} = 0.01$ dm s$^{-1}$; $k_{off} = 0.1$ s$^{-1}$, (B) GFP-Ras2 data with model predictions for diffusion and fast exchange with parameters $D = 0.0691$ μm$^2$/s; $\alpha = 27.7$ s$^{-1}$; $k_{on} = 0.1$ dm s$^{-1}$; $k_{off} = 1$ s$^{-1}$, (C) GFP-Ras2C318S data with model predictions for diffusion and slow exchange with parameters $D = 0.2743$ μm$^2$/s; $\alpha = 43.09$ s$^{-1}$; $k_{on} = 0.01$ dm s$^{-1}$; $k_{off} = 0.1$ s$^{-1}$, (D) GFP-Ras2C318S data with model predictions for diffusion and fast exchange with parameters $D = 0.2768$ μm$^2$/s; $\alpha = 43.5$ s$^{-1}$; $k_{on} = 0.1$ dm s$^{-1}$; $k_{off} = 1$ s$^{-1}$

6. FRAP data summary for GFP-Ras2 and GFP-PMA1 for 1μm x 1μm bleach ROI. The integral membrane protein GFP-PMA1 diffuses much slower when compared to lipid anchored GFP-Ras2.

7. Averaged bleach horizontal bleach profile data for a GFP-Ras2 1μm x 1μm bleach ROI experiment calculated from the bleach frame shown in Fig 2B, and model prediction using the estimated parameters $D = 0.0691$ μm$^2$/s, $\alpha = 27.7$ s$^{-1}$.

Figure 1

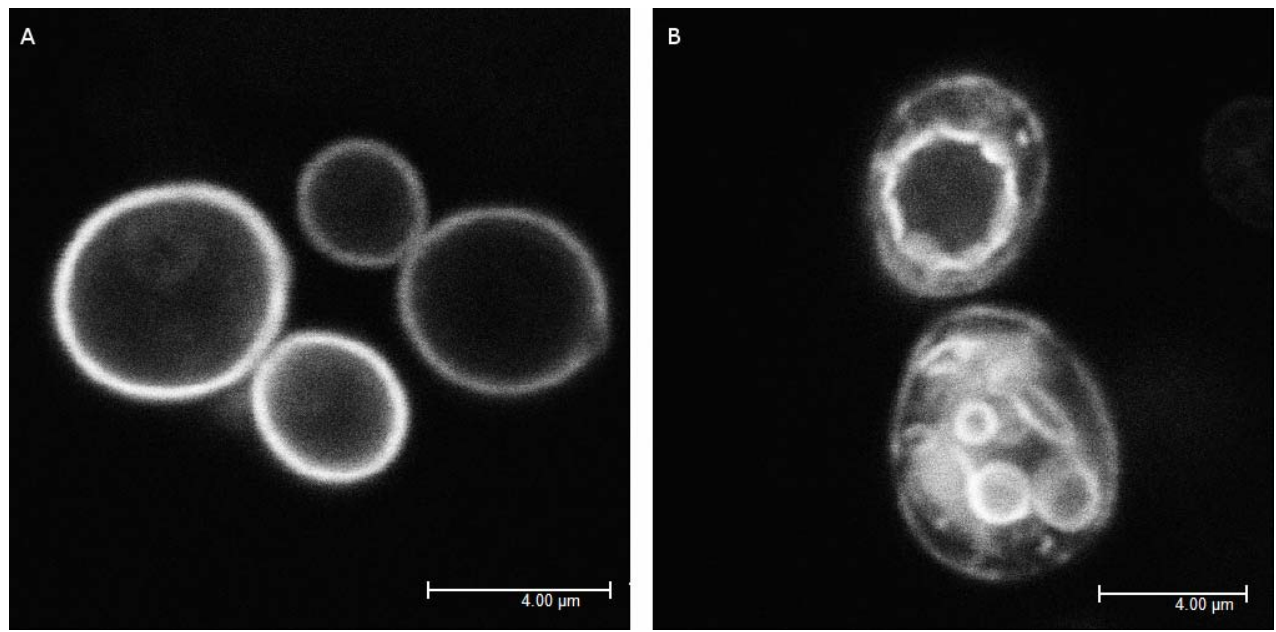

Figure 2

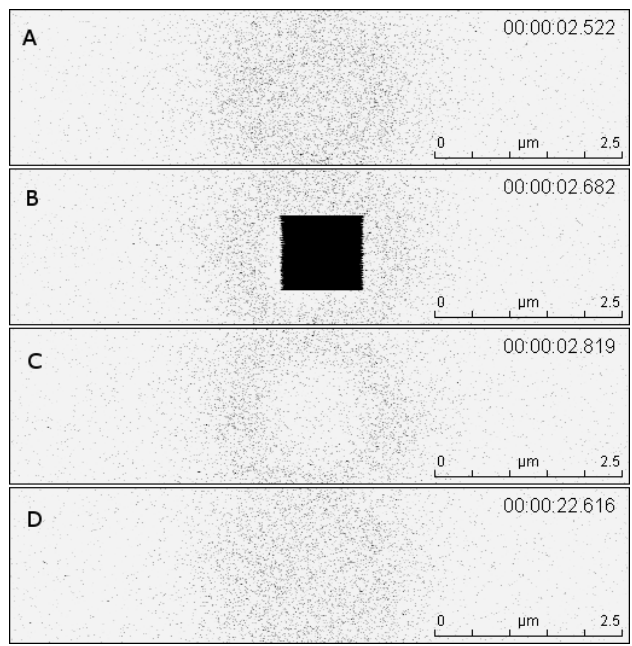

Figure 3

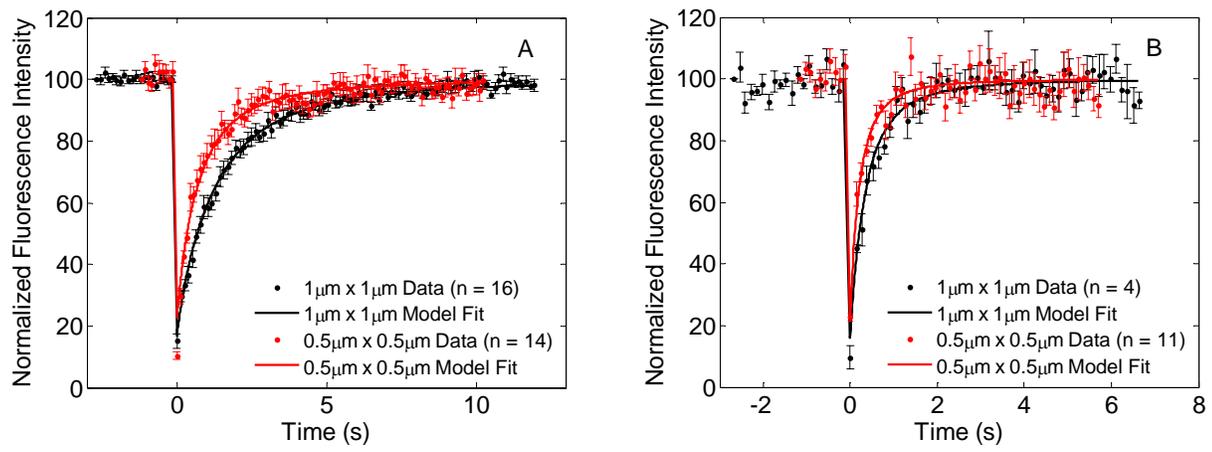

Figure 4

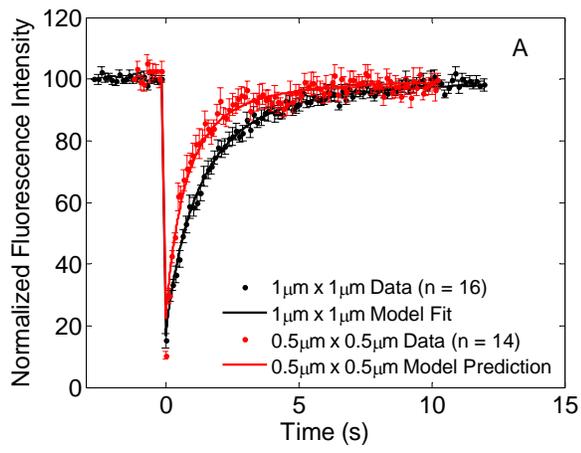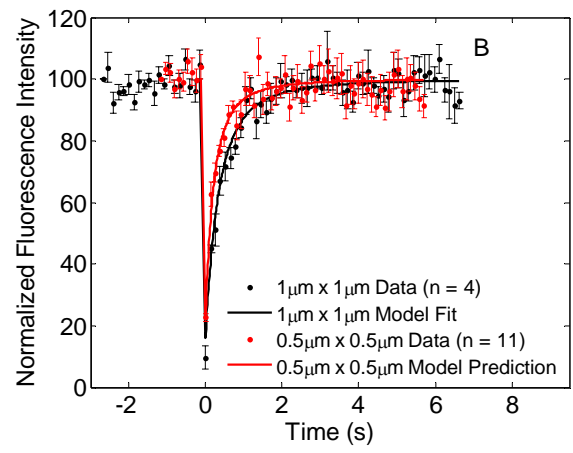

Figure 5

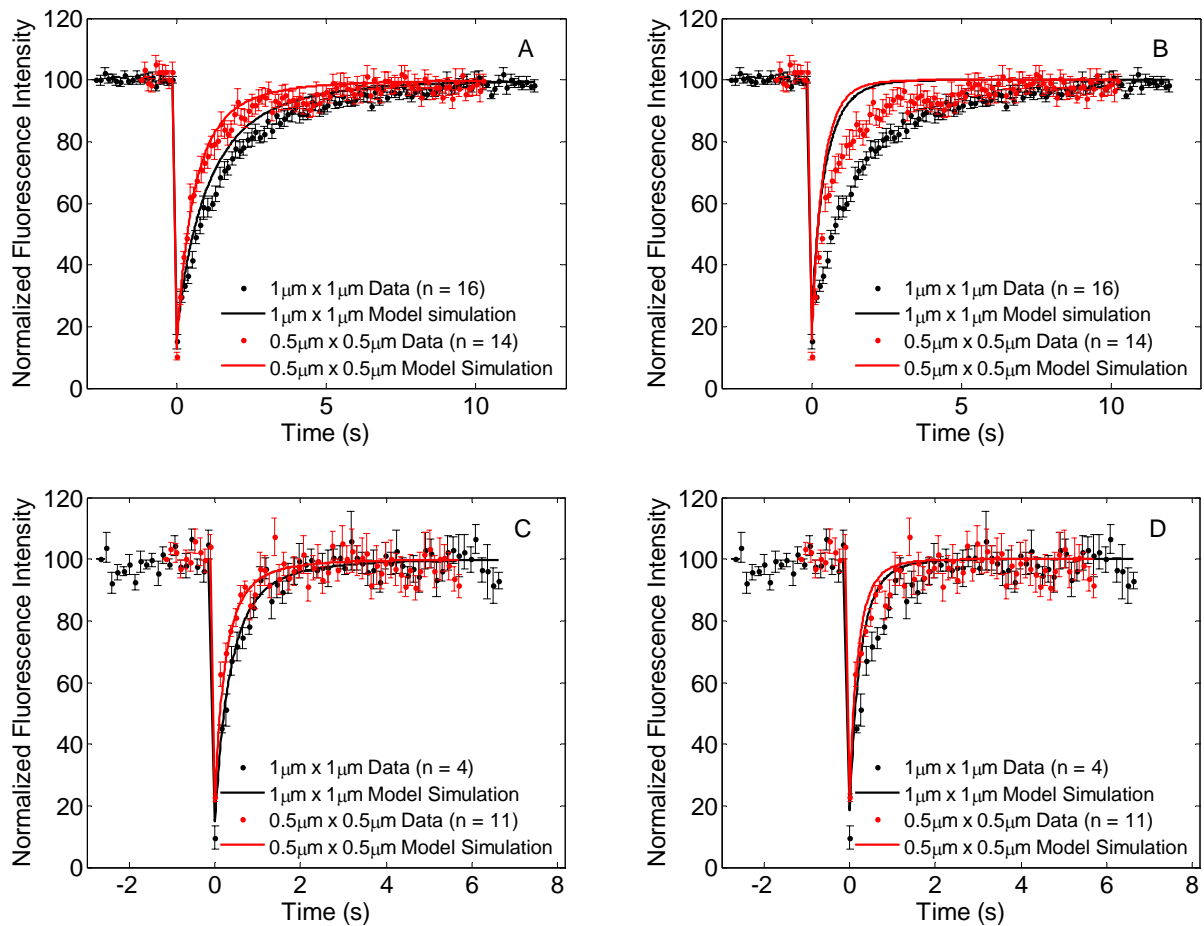

Figure 6

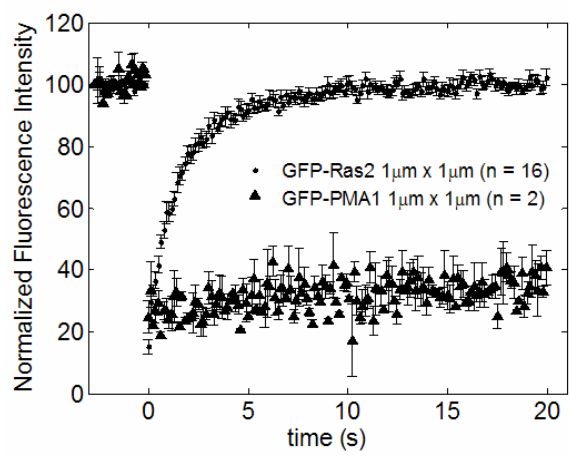

Figure 7

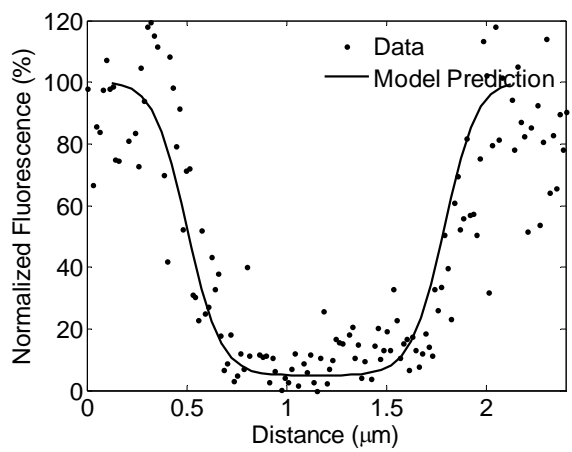